\documentclass{llncs}

\usepackage{hyperref}
\usepackage{tikz}
\usetikzlibrary{decorations.markings}
\usetikzlibrary{arrows}
\usepackage{amssymb}
\usepackage{amsfonts,amsmath,latexsym,stmaryrd}

\usepackage{listings}

\newcommand{\C}[1]{\mbox{\lstinline`#1`}}

\definecolor{dkblue}{rgb}{0,0.1,0.5} 
\definecolor{lightblue}{rgb}{0,0.5,0.5} 
\definecolor{dkgreen}{rgb}{0,0.4,0} 
\definecolor{dk2green}{rgb}{0.4,0,0} 
\definecolor{dkviolet}{rgb}{0.6,0,0.8}
\definecolor{shadethmcolor}{rgb}{0.9, 0.9,1}

\usepackage{rotating}
\usepackage{pdflscape}

\usepackage{theorem}

\theorembodyfont{\rmfamily}
\newtheorem{Thm}{Theorem}

\newtheorem{Lem}[Thm]{Lemma}
\newtheorem{PS}{Proof Strategy}
\newtheorem{PR}{Problem}

\newtheorem{Alg}{Algorithm}

\begin{document}

\title{ML4PG in Computer Algebra Verification\thanks{The work was supported by EPSRC grant EP/J014222/1. The final publication is available at \url{http://link.springer.com.}}}

\author{J\'onathan Heras \and Ekaterina Komendantskaya}
\authorrunning{J. Heras and E. Komendantskaya}

\institute{School of Computing, University of Dundee, UK\\
\email{\{jonathanheras,katya\}@computing.dundee.ac.uk}}

\maketitle

\begin{abstract}

ML4PG is a machine-learning extension that provides statistical proof hints during the process of 
Coq/SSReflect proof development. In this paper, we use ML4PG to find proof patterns in the CoqEAL 
library -- a library that was devised to verify the correctness of Computer Algebra algorithms. 
In particular, we use ML4PG to help us in the formalisation of an efficient algorithm to compute
the inverse of triangular matrices.\\
\textbf{Keywords:} ML4PG, Interactive Theorem Proving, Coq, SSReflect, Machine Learning, Clustering, CoqEAL.

\end{abstract}

\section{Introduction}

There is a trend in interactive theorem provers to develop general purpose methodologies to aid in the
formalisation of a family of related proofs. However, although the application of a methodology is 
straightforward for its developers, it is usually difficult for an external user to decipher the key 
results to import such a methodology into a new development. Therefore, tools which can capture methods
and suggest appropriate lemmas based on proof patterns would be valuable. ML4PG~\cite{HK12}
-- a machine-learning extension to Proof General that interactively finds proof patterns in Coq/SSReflect -- 
can be useful in this context. 

In this paper, we use ML4PG to guide us in the formalisation of a fast algorithm to compute the inverse of
triangular matrices using the CoqEAL methodology~\cite{DMS12} -- a method designed to verify the correctness of 
efficient Computer Algebra algorithms.

\emph{Availability.} ML4PG is accessible from~\cite{HK12}, where the reader can find related papers,
examples, the links to download ML4PG and all libraries and proofs we mention here. 

\section{Combining the CoqEAL methodology with ML4PG}

Most algorithms in modern Computer Algebra systems are designed to be efficient, and 
this usually means that their verification is not an easy task. In order to overcome this 
problem, a methodology based on the idea of \emph{refinements} was presented in~\cite{DMS12},
and was implemented as a new library, built on top of the SSReflect libraries, called 
\emph{CoqEAL}. The approach~\cite{DMS12}  to formalise efficient
algorithms can be split into three steps: 
\begin{itemize}
\item[\textbf{S1.}] define the algorithm relying on rich dependent 
types, as this will make the proof of its correctness easier; 
\item[\textbf{S2.}] refine this definition to an
efficient algorithm described on high-level data structures; and, 
\item[\textbf{S3.}] implement it on data structures
which are closer to machine representations. 
\end{itemize}

The CoqEAL methodology is clear and the authors have shown that it can be 
extrapolated to different problems. Nevertheless, this library contains 
approximately 400 definitions and 700 lemmas; and the search of proof strategies inside this library
is not a simple task if undertaken manually. Intelligent proof-pattern recognition methods could help with 
such a task. 

In order to show this, let us consider the formalisation of a fast algorithm to
compute the inverse of triangular matrices over a field with $1$s in the diagonal using
the CoqEAL methodology. SSReflect already implements the matrix inverse relying on rich 
dependent types using the \lstinline?invmx? function; then, we only need to focus on the
second and third steps of the CoqEAL methodology. We start defining a function called 
\lstinline?fast_invmx? using high-level data structures. 

\begin{Alg}

Let $M$ be a square triangular matrix of size $n$ with $1$s in the diagonal; then \lstinline?fast_invmx(M)? is recursively defined
as follows. 

\begin{itemize}
 \item If $n=0$, then \lstinline?fast_invmx(M)=1%M? (where \lstinline?1%M? is the notation for the identity matrix in SSReflect).
 \item Otherwise, decompose $M$ in a matrix with four components: the top-left element, which is $1$; the top-right line vector,
 which is null; the bottom-left column vector $C$; and the bottom-right $(n-1)\times (n-1)$ matrix $N$; that is,  
 $M=\left(\begin{array}{c|c} 
1 & 0\\ 
 \hline
 C & N
   \end{array}\right)$. Then define \lstinline?fast_invmx(M)? as:
    \begin{center}
     
\lstinline?fast_invmx(M)?$=\left(\begin{array}{c|c} 
1 & 0\\ 
 \hline
 - \verb"fast_invmx(N) *m C" & \verb"fast_invmx(N)"
   \end{array}\right)$
    \end{center}
    
 \noindent where \lstinline?*m? is the notation for matrix multiplication in SSReflect.
\end{itemize}

\end{Alg}

Subsequently, we should prove the equivalence between the functions \lstinline?invmx?
and \lstinline?fast_invmx? -- Step S2 of the CoqEAL methodology. Once this result is proven,
we can focus on the third step of the CoqEAL methodology. It is worth mentioning that neither
\lstinline?invmx? nor \lstinline?fast_invmx? can be used to actually compute the inverse of matrices.
These functions cannot be executed since the definition of matrices is locked in SSReflect to avoid
the trigger of heavy computations during deduction steps. Using Step S3 of the CoqEAL methodology, 
we can overcome this pitfall. In our case, we implement the function \lstinline?cfast_invmx? using
lists of lists as the low level data type for representing matrices and to finish the formalisation
we should prove the following lemma. 

\begin{Lem}\label{lem}
Let $M$ be a square triangular matrix of size $n$ with $1$s in the diagonal; then given $M$ as input, 
\lstinline?fast_invmx? and \lstinline?cfast_invmx? obtain the same result but with different representations.
The statement of this lemma in SSReflect is: 

\begin{lstlisting}
 Lemma cfast_invmxP : forall (n : nat) (M : 'M_n),
    seqmx_of_mx (fast_invmx M) = cfast_invmx (seqmx_of_mx M). 
\end{lstlisting}

\noindent where the function \lstinline?seqmx_of_mx? transforms matrices represented as functions to 
matrices represented as lists of lists. 

\end{Lem}

The proof of Lemma~\ref{lem} for a non-expert user of CoqEAL is not direct, and, after applying induction
on the size of the matrix, the developer can get easily stuck when proving such a result. 

\begin{PR}
 Find a method to proceed with the inductive case of Lemma~\ref{lem}.
\end{PR}

\begin{figure}
 \centering
  \includegraphics[height=3cm,width=10cm]{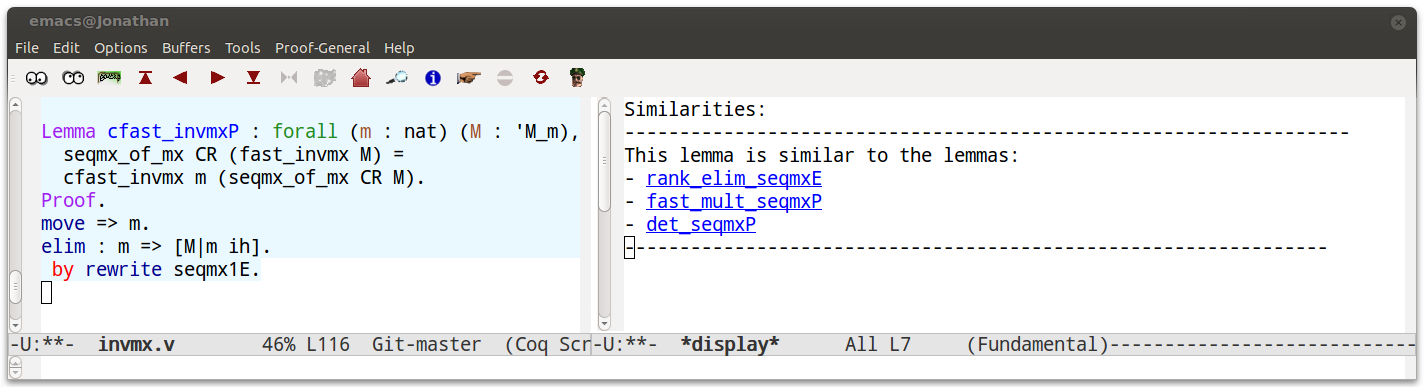}
  \caption{Suggestions for Lemma \lstinline?cfast_invmxP?. The Proof General window has been split into two windows positioned side by
side: the left one keeps the current proof script, and the right one shows the suggestions 
provided by ML4PG.}\label{screenshot}
 \end{figure}

In this context, the user can invoke ML4PG to find some common proof-pattern in the CoqEAL library. 
ML4PG generated solutions is presented in Figure~\ref{screenshot}.

ML4PG suggests three lemmas which are the equivalent counterparts of Lemma~\ref{lem} for the algorithms computing the rank, the determinant and
the fast multiplication of matrices. Inspecting the proof of these three lemmas, the user can find Proof Strategy~\ref{ps} which is followed 
by those three lemmas and which can also be applied in Lemma~\ref{lem}.

\begin{PS}\label{ps}
 Apply the \emph{morphism lemma} to change the representation from abstract matrices to executable ones.
 Subsequently, apply the \emph{translation lemmas} of the operations involved in the algorithm -- translation
 lemmas are results which state the equivalence between the executable and the abstract counterparts of several operations
 related to matrices.
\end{PS}

It is worth remarking that the user is left with the task of finding a proof strategy from the suggestions 
provided by ML4PG. In the future, we could apply symbolic machine-learning techniques such as Rippling~\cite{BB05}
and Theory Exploration~\cite{CJRS13} to automatically conceptualise the proof strategies from the suggestions provided by ML4PG. 

\section{Applying ML4PG to the CoqEAL library}

In the section, we show how ML4PG discovers the lemmas which follow Proof Strategy~\ref{ps}. This process can be 
split into 4 steps: extraction of significant features from library-lemmas, selection of the machine-learning algorithm, 
configuration of parameters, and presentation of the output. 

\textbf{Step 1. Feature extraction.}  During the proof development, ML4PG works on the background of Proof General,
 and extracts (using the algorithm described in~\cite{HK12}) some simple, low-level features from interactive proofs 
 in Coq/SSReflect. In addition, ML4PG extends Coq's compilation procedure to extract lemma-features from
 already-developed libraries. 
 
 In the example presented in the previous section, we have extracted the features from the $18$ files included in the
 CoqEAL library (these files involve 720 lemmas). Any number of additional Coq libraries can be be selected using the ML4PG menu. 
Unlike e.g. \cite{Mash}, scaling is done at the feature extraction stage, rather than on the machine-learning stage of the process. 

\textbf{Step 2. Clustering algorithm.} On user's request, ML4PG sends the gathered statistics to a chosen machine-learning
 interface and triggers execution of a clustering algorithm of the user's choice -- clustering algorithms~\cite{Bishop} 
 are a family of unsupervised learning methods which divide data into $n$ groups of similar objects (called clusters),
 where the value of $n$ is provided by the user.

 We have integrated ML4PG with several clustering algorithms available in MATLAB (K-means and Gaussian) and Weka 
 (K-means, FarthestFirst and Expectation Maximisation). In the CoqEAL example, ML4PG uses the MATLAB K-means algorithm 
 to compute clusters -- this is the algorithm used by default.

\textbf{Step 3. Configuration of granularity.} The input of the clustering algorithms is a file that contains the 
information associated with the lemmas to be analysed, and a natural number $n$, which indicates the number of clusters. 
The file with the features of the library-lemmas is automatically extracted (see~\cite{HK12}).

To determine the value 
of $n$, ML4PG has its own algorithm that calculates the optimal number of clusters interactively, based on the library
size. As a result, the user does not provide the value of $n$ directly, but just decides on granularity in the ML4PG menu.
The granularity parameter ranges from $1$ to $5$, where $1$ stands for a low granularity (producing a few large clusters
with a low correlation among their elements) and $5$ stands for a high granularity (producing many smaller clusters with
a high correlation among their elements). By default, ML4PG works with the granularity value of $3$ and this is the value
presented in the previous section. 

\textbf{Step 4. Presentation of the results.} Clustering algorithms output contains not only clusters but also a measure
which indicates the proximity of the elements of the clusters. In addition, results of one run of a clustering algorithm
may differ from another; then ML4PG runs the clustering algorithm 200 times, obtaining the frequency of each cluster as a
result. These two measures (proximity and frequencies) are used as thresholds to decide 
on the single ``most reliable'' cluster to be shown to the
user, cf.~Figure~\ref{screenshot}.

These 4 steps are the workflow followed by ML4PG to obtain clusters of similar proofs. Let us present now the results 
that ML4PG will obtain if the user varies the different parameters -- these results are summarised in Table~\ref{tab:compare}. 

\begin{table}
\centering
  \begin{tabular}{|c|c|c|c|c|c|c|c|c|c|c|c|c|c|c|c|}
   \hline
	    & $g=1$ & $g=2$&$g=3$& $g=4$&$g=5$\\
     Algorithm: & ($n=72$)& ($n=80$)& ($n=90$)& ($n=102$)& ($n=120$)\\
	\hline
   Gaussian &$24^{a,b,c,d}$ & $12^{a,b,c,d}$ & $10^{a,b,c,d}$& $10^{a,b,c,d}$& $10^{a,b,c,d}$ \\
   \hline
   K-means (Matlab) &$20^{a,b,c,d}$ & $14^{a,b,c,d}$ & \textbf{4}$^{a,b,c,d}$& $0$& $0$ \\
   \hline
   K-means (Weka) & $16^{a,b,c,d}$& $11^{a,b,c,d}$  & \textbf{4}$^{a,b,c,d}$ & $0$ & $0$  \\
   \hline
   Expectation Maximisation  &$52^{a,b,c,d}$ &$45^{a,b,c,d}$  &$43^{a,b,c,d}$ &$39^{a,b,c,d}$ &  $14^{a,b,c,d}$\\
   \hline
   FarthestFirst  &$30^{a,b,c,d}$ &$27^{a,b,c,d}$  &$27^{a,b,c,d}$ &$26^{a,b,c,d}$ & $20^{a,b,c,d}$ \\
   \hline
     \end{tabular} 
     
  \caption{\footnotesize{\textbf{A series of clustering experiments discovering Proof Strategy~\ref{ps}.} The table shows the sized of clusters
  containing: a) Lemma  \lstinline?cfast_invmxP?, b) Lemma about rank (\lstinline?rank_elim_seqmxE?), c) Lemma about fast multiplication 
  (\lstinline?fast_mult_seqmxP?), and d) Lemma about determinant (\lstinline?det_seqmxP?).}}\label{tab:compare}
 \end{table}
 
As can be seen in Table~\ref{tab:compare}, the clusters obtained by almost all variations of the learning algorithms and parameters include the 
lemmas which led us to formulate Proof Strategy~\ref{ps}. However, there are some remarkable differences among the results. First of all, the results
obtained with the Expectation Maximisation and FarthestFirst algorithms include several additional lemmas that make difficult the discovery of
a common pattern. The same happens with the other algorithms for granularity values $1$ and $2$; however the 
clusters can be refined when increasing the granularity value. The results are clusters of a sensible size which contain lemmas with a high correlation; 
allowing us to spot Proof Strategy~\ref{ps}.

\bibliographystyle{plain}
\bibliography{cicm2013}

\begin{thebibliography}{1}

\bibitem{BB05}
D.~Basin, A.~Bundy, D.~Hutter, and A.~Ireland.
\newblock {\em Rippling: Meta-level Guidance for Mathematical Reasoning}.
\newblock Cambridge University Press, 2005.

\bibitem{Bishop}
C.~Bishop.
\newblock {\em Pattern Recognition and Machine Learning}.
\newblock Springer, 2006.

\bibitem{CJRS13}
K.~Claessen, M.~Johansson, D.~Ros\'en, and N.~Smallbone.
\newblock Automating inductive proofs using theory exploration.
\newblock In {\em Proceedings 24th International Conference on Automated
  Deduction (CADE-24)}, LNCS, 2013.

\bibitem{DMS12}
M.~D\'en\`es, A.~M\"ortberg, and V.~Siles.
\newblock {A Refinement Based Approach to Computational Algebra in Coq}.
\newblock In {\em Proceedings 3rd Conference on Interactive Theorem Proving
  (ITP'12)}, volume 7406 of {\em LNCS}, pages 83--98, 2012.

\bibitem{HK12}
J.~Heras and E.~Komendantskaya.
\newblock {ML4PG: downloadable programs, manual, examples}, 2012--2013.
\newblock \url{www.computing.dundee.ac.uk/staff/katya/ML4PG/}.

\bibitem{Mash}
D.~K\"uhlwein, J.~C. Blanchette, C.~Kaliszyk, and J.~Urban.
\newblock {MaSh: Machine Learning for Sledgehammer}.
\newblock In {\em Proceedings of ITP'13}, LNCS, 2013.

\end{thebibliography}

\end{document}